\documentclass[superscriptaddress,aps,prb,nofootinbib,amsmath,amssymb,reprint]{revtex4-1}
\usepackage[utf8]{inputenc}
\usepackage{graphicx}
\usepackage{amsmath,amssymb}
\usepackage{textcomp}
\usepackage{color}
\usepackage{xcolor}
\usepackage{epstopdf}
\usepackage{siunitx}
\usepackage{mathtools}
\usepackage{natbib}
\usepackage[normalem]{ulem}
\usepackage{soul}

\usepackage{extarrows}
\usepackage{calrsfs}

\newcommand{\CCO}{Ca$_3$Co$_2$O$_6$ }

\newcommand{\UDD}{$\uparrow \downarrow \downarrow$ }
\newcommand{\UUD}{$\uparrow \uparrow \downarrow$ }
\newcommand{\UUU}{$\uparrow \uparrow \uparrow$ }

\newcommand{\real}{$\chi '$ }
\newcommand{\imag}{$\chi ''$ }

\begin{document}

\title{Peculiar magnetic dynamics across the in-field transition in \CCO}

\author{Nagabhushan G.Hegde}
\affiliation{Laboratory for Quantum Magnetism, Institute of Physics, Ecole Polytechnique Federale de Lausanne, CH-1015 Lausanne, Switzerland}
\author{Ivana Levati\'c}
\affiliation{Institute of Physics, Bijeni\v cka 46, HR-10 000, Zagreb, Croatia}
\author{Arnaud Magrez}
\affiliation{Crystal Growth Facility, Institute of Physics, Ecole Polytechnique Federale de Lausanne, CH-1015 Lausanne, Switzerland}
\author{Henrik M. R\o{}nnow}
\affiliation{Laboratory for Quantum Magnetism, Institute of Physics, Ecole Polytechnique Federale de Lausanne, CH-1015 Lausanne, Switzerland}
\author{Ivica \v{Z}ivkovi\'{c}}
\email{ivica.zivkovic@epfl.ch}
\affiliation{Laboratory for Quantum Magnetism, Institute of Physics, Ecole Polytechnique Federale de Lausanne, CH-1015 Lausanne, Switzerland}

\begin{abstract}
The discovery of multiple coexisting magnetic phases in a crystallographically homogeneous compound Ca$_3$Co$_2$O$_6$ has stimulated an ongoing research activity. In recent years the main focus has been on the zero field state properties, where exceedingly long time scales have been established. In this study we report a detailed investigation of static and dynamic properties of Ca$_3$Co$_2$O$_6$ across the magnetic field induced transition around 3.5 T. This region has so far been practically neglected while we argue that in some aspects it represents a simpler version of the transition across the $B = 0$ state. Investigating the frequency dependence of the ac susceptibility we reveal that on the high field side ($B > 3.5$ T) the response corresponds to a relatively narrow distribution of magnetic clusters. The distribution appears very weakly dependent on magnetic field, with an associated energy barrier of around 200 K. Below 3.5 T a second contribution arises, with much smaller characteristic frequencies and a strong temperature and magnetic field dependence. We discuss these findings in the context of intra-chain and inter-chain clustering of magnetic moments.
\end{abstract}

\date{\today} 
\maketitle

\section{Introduction}

Magnetic compounds that exhibit geometrical frustration, coupled with low dimensional characteristics, provide one of the most exciting playgrounds for discovery of novel phases with potentially exotic properties. Prime examples are quantum spin liquids~\cite{Zhou2017}, where both of these features play a crucial role in preventing the system from attaining long-range order (LRO).

On the other hand there are more classical systems where several configurations of magnetic moments are nearly degenerate, creating a landscape of metastable states with slow dynamics, whose time scale strongly depends on external parameters like temperature and magnetic field. Given that typical examples of slow dynamics are associated with disordered spin-glasses and inherently heterogeneous super-paramagnetic systems~\cite{Mydosh1993}, the occurrence of metastable states in homogeneous magnetic systems attracts considerable attention.~\cite{Paulsen2014}

\CCO has been in focus since the first discovery of magnetization plateaus~\cite{Aasland1997}, with suggestions of involvement of quantum tunnelling processes due to the similarity to single-molecule magnets.~\cite{Hardy2004a} Further investigations showed a complex time-, temperature- and magnetic field-history dependence of the observed plateaus but more intriguingly the same has been established for its zero-field state. With Ising-type magnetic moments residing on cobalt ions~\cite{Leedahl2019}, the magnetic lattice consists of quasi-one-dimensional chains running along the $c$ axis arranged in a triangular pattern in the $ab$ plane. There is a dominant, ferromagnetic (FM) intra-chain interaction $J_1 \sim 24$ K, while the inter-chain interactions are antiferromagnetic (AFM), more than order of magnitude weaker.~\cite{Allodi2014} Due to the relative shift of neighbouring chains by $\pm c/3$, there is a finite effect of the inter-chain coupling along the $c$ axis and a consequent competition with the dominant $J_1$ interaction, which profoundly influences the properties of this compound.

It has proven to be very difficult to accurately describe the zero field properties of \CCO. Neutron diffraction experiments indicated that below $T_N = 25$ K an incommensurate, amplitude-modulated spin-density wave (SDW) is formed~\cite{Agrestini2008}, whose periodicity strongly depends on temperature and magnetic field~\cite{Moyoshi2011,Motoya2018}. A surprising result was the reduction of magnetic Bragg peak intensities with further cooling, which was ascribed later to short-range order (SRO) development.~\cite{Petrenko2005,Agrestini2011,Prsa2014}. A second LRO phase has been discovered below 13 K, with a wave vector {\boldmath $k$} $ = (0.5,-0.5,1)$ and an exceptionally long time scales but involving only up to 20\% of sample's volume fraction.~\cite{Agrestini2011} Moyoshi and Motoya~\cite{Moyoshi2011} argued that the actual modulated structure is of a square-wave type while quantum Monte Carlo simulations indicate~\cite{Kamiya2012} a formation of a quasi-periodic soliton lattice, where the mean separation between solitons determines the periodicity of the observed density wave.

\begin{figure}[b]
	\begin{center}
		\includegraphics[width=0.9\columnwidth]{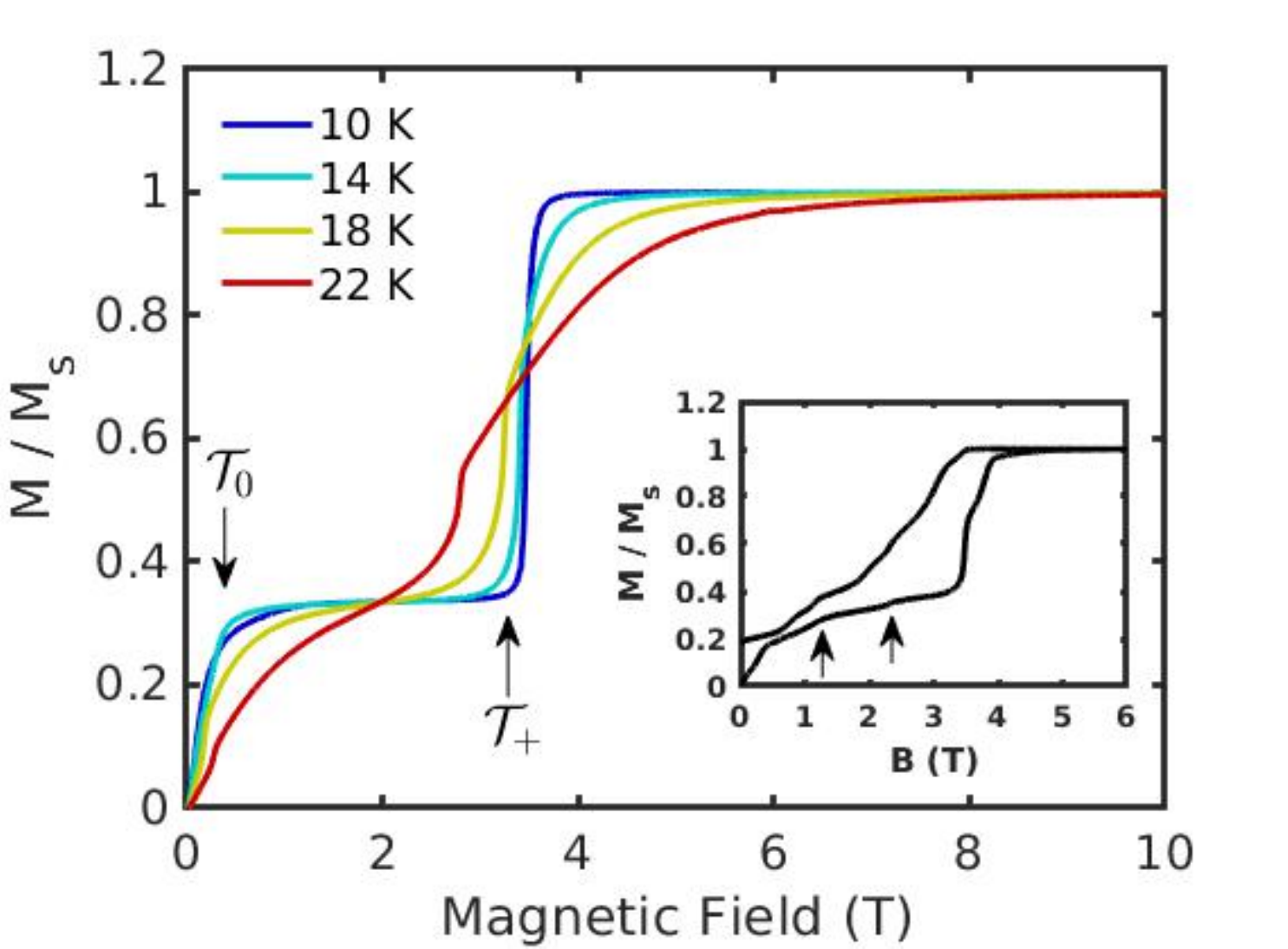}
		\caption{Magnetic field dependence of magnetization in \CCO. Inset: $T = 2$ K with significant hysteretic effects and additional steps that are highlighted with arrows.}
		\label{fig::Hdependence}
	\end{center}
\end{figure}

To grasp the complexity of the ground state and the source of its highly dynamic features one needs to explore the system away from the $B = 0$ condition. As shown in Figure~\ref{fig::Hdependence} between 10 K and 15 K well defined steps of magnetization can be observed at 1/3$M_s$ and at full saturation $M_s$, where the value of $M_s = 5.2 \mu_B$, in good agreement with neutron scattering results.~\cite{Moyoshi2011} There are four steps in total (including -$M_s$ and -1/3$M_s$) with three transitions between them:

\begin{equation}
\label{eq::Htransitions}
\downarrow \downarrow \downarrow \; \xlongleftrightarrow{\mathcal{T}_-} \; \uparrow \downarrow \downarrow \; \xlongleftrightarrow{\mathcal{T}_0} \; \uparrow \uparrow \downarrow \; \xlongleftrightarrow{\mathcal{T}_+} \; \uparrow \uparrow \uparrow
\end{equation}

\UDD and \UUD configurations represent an average of a large assembly of $c$ axis chains so that local deviations are averaged out. Also, from the perspective of symmetry, one can consider $\mathcal{T}_-$ and $\mathcal{T}_+$ equivalent so for further discussion we will focus on $\mathcal{T}_0$ and $\mathcal{T}_+$.

If one looks back to the $B = 0$ state, it is realized that it is at a particular point in the phase diagram, exactly in the middle of $\mathcal{T}_0$. When magnetic field is ramped from the -1/3$M_s$ plateau across $B = 0$, it initiates a system-wide spin reversal from \UDD to \UUD state, where one spin out of three changes its orientation. However, the exact development of this reversal has never been explored. It is evident that the reversal does not occur suddenly, i.e. we do not see a simultaneous switch of all the spins at once (compare this to very sharp transitions in single-molecule magnets with avalanche processes occurring during the reversal~\cite{Decelle2009}). Thus, only small segments of each chain switch direction one at a time. On a pictorial level the \UUD state emerges within the \UDD state, first as sparsely distributed small droplets, which then grow and coalesce as the field is further increased, and then finally the system reaches the 1/3$M_s$ plateau. The exactly symmetrical situation occurs on the other side of $\mathcal{T}_0$ when the field is reversed, where small droplets of the \UDD state exist and coalesce within an otherwise homogeneous \UUD state. This description is very similar to general concepts applied in first-order phase transitions~\cite{Binder1987}, where two phases coexist, with latent heat being released or absorbed depending on the direction of the process.

This pictorial representation agrees with the proposition that magnetic microphases~\cite{Kamiya2012} exist in \CCO, with their sizes being spread over a finite distribution. In such a case, the zero field state can be envisaged as a highly dynamic ensemble where both \UDD and \UUD phases coexist on all spatial scales and where the exact distribution of cluster sizes of both phases is highly sensitive to the exact history path through the temperature-magnetic field phase space. Measuring the dynamic properties are also compounded by the fact that the two phases contain both orientations of spins ($up$ and $down$) so it becomes practically impossible to distinguish between the dynamic and static part or even to clearly mark the boundaries of phases.

On the other hand, within the $\mathcal{T}_+$ transition it is clear that the dynamic part is carried by segments of $\downarrow$ spins, while the static background contains only $\uparrow$ spins. From this perspective, $\mathcal{T}_+$ represents a somewhat simpler version of $\mathcal{T}_0$. This reasoning led us to investigate closely the frequency, temperature and magnetic field dependence of $\mathcal{T}_+$.

\section{Experimental details}

\CCO crystals were grown by the flux method. Single phase Ca$_3$Co$_4$O$_9$ powder was first prepared by annealing a stoichiometric mixture of CaCO$_3$ and Co$_3$O$_4$ at 1000$^0$C for three days. The powder was then mixed with K$_2$CO$_3$ used as flux in a 1/7 weight ratio. The mixture was annealed at 1050$^0$C for 12 hours. The melt was then cooled down to 850$^0$C at a rate of 1$^0$C/h and then to room temperature at a rate of 10$^0$C/h. The flux was dissolved in distilled water. Dark green crystals of \CCO, with the longest direction parallel to the $c$ axis could be collected.

Magnetization and ac susceptibility measurements were performed using a commercial Quantum Design PPMS. For magnetization we used a VSM mode with sample cooled from 30 K ($> T_N = 25$ K) to the desired temperature in zero field with a subsequent field sweep to 14 T using 100 Oe/s. For ac susceptibility measurements we used an induction method, with the field scan limited from 3.3 T to 3.7 T, encompassing the majority of the $\mathcal{T}_+$ transition. The amplitude of the ac magnetic field was set to 5 Oe for all frequencies in the range 10 Hz -- 10 kHz.

\begin{figure}[b]
	\begin{center}
		\includegraphics[width=0.9\columnwidth]{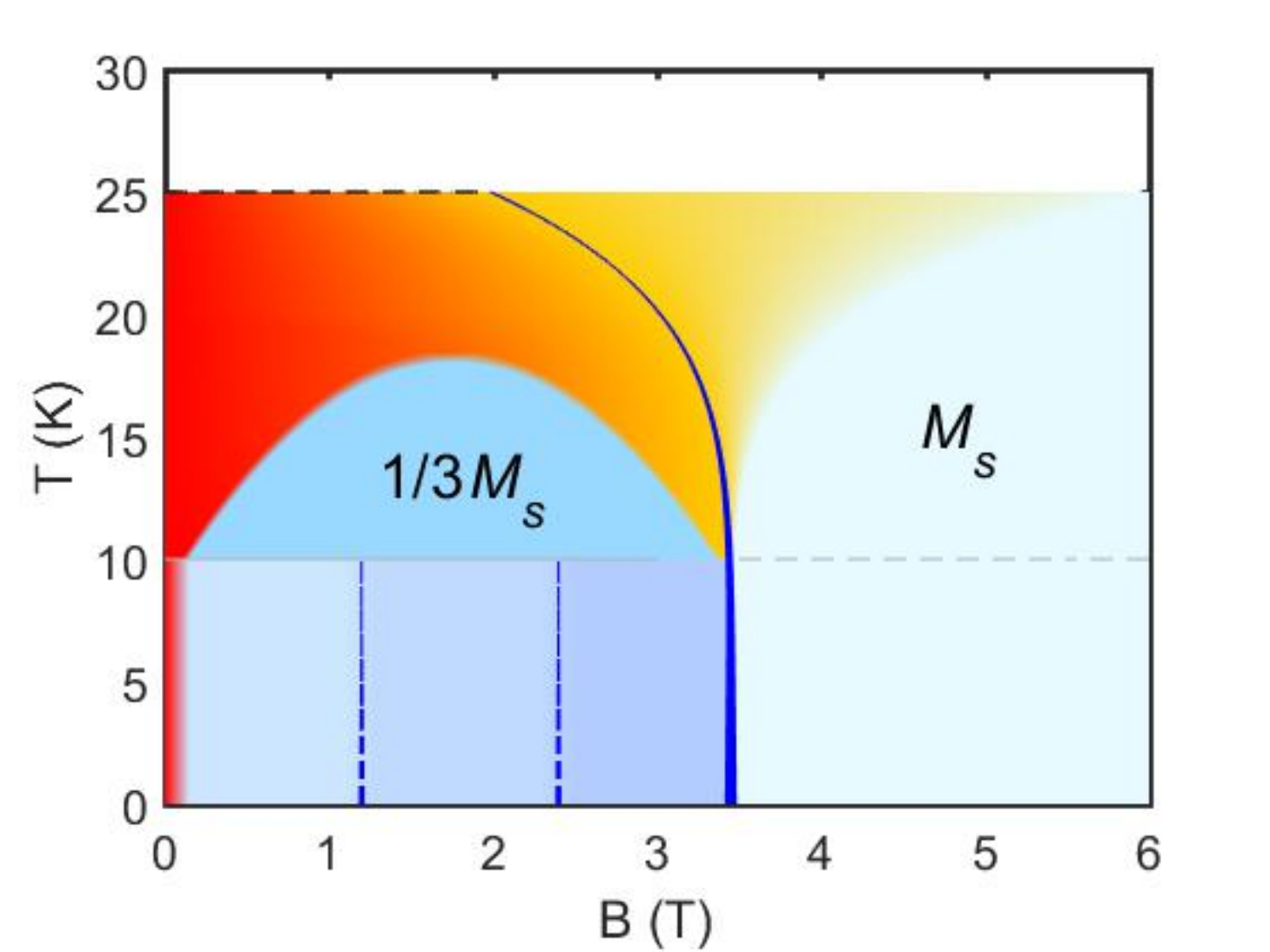}
		\caption{Temperature -- magnetic field phase diagram of \CCO. The plateaus at 1/3$M_s$ and $M_s$ are approximately shaded. Below the 10 K line hysteretic behaviour emerges.}
		\label{fig::phasediagram}
	\end{center}
\end{figure}

\section{Experimental results}

In Figure~\ref{fig::Hdependence} we present the magnetization data. Around 10 K \CCO exhibits well defined plateaus. The plateaus are smeared out as temperature is raised towards the transition temperature $T_N$ and the jumps become less sharp. On the other hand, lowering temperature below 10 K induces a development of hysteretic effects and additional steps can be seen, marked by arrows in the inset of Figure~\ref{fig::Hdependence}.

\begin{figure}[t]
	\begin{center}
		\includegraphics[width=0.9\columnwidth]{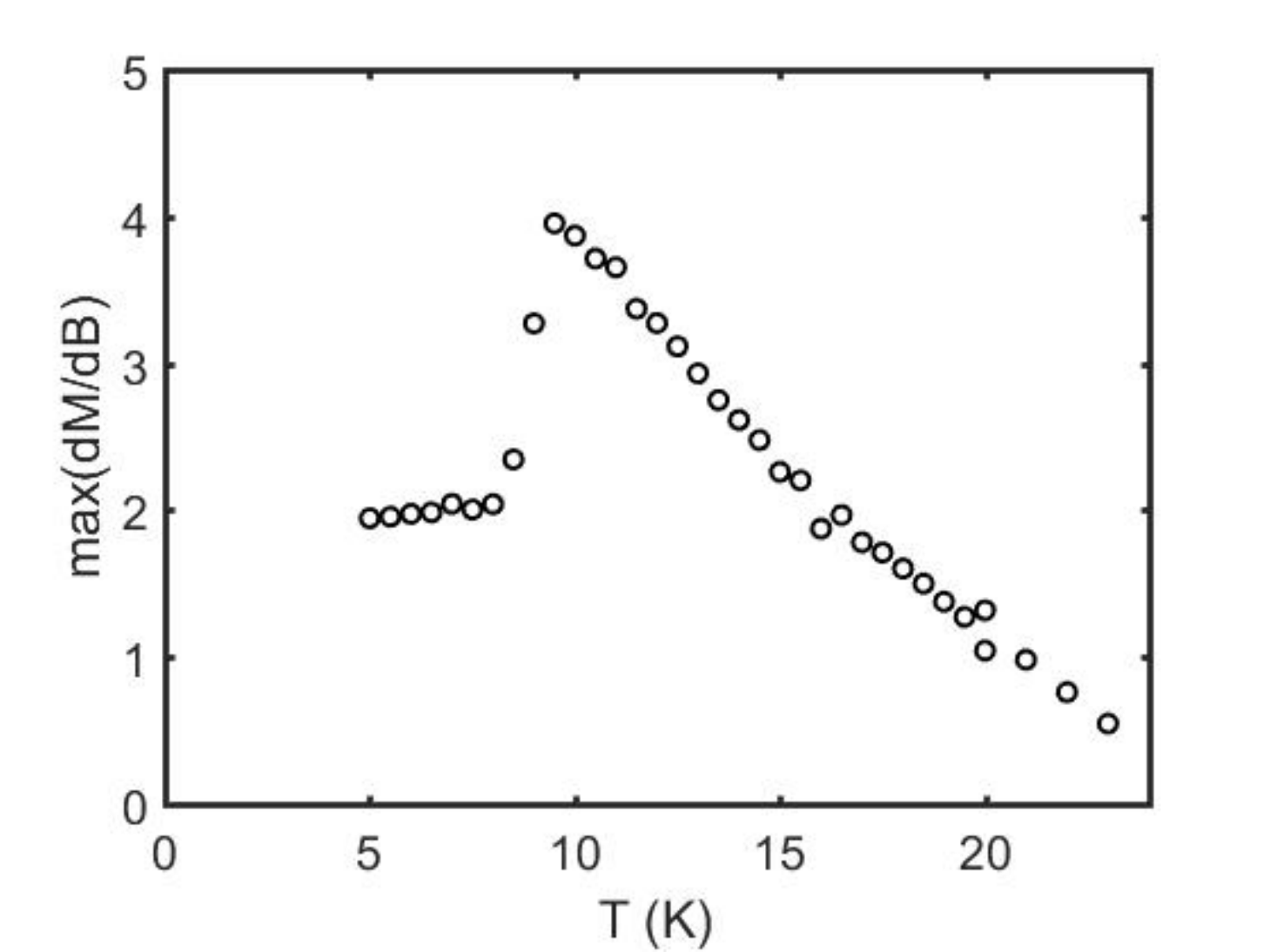}
		\caption{Temperature dependence of the maximum value of derivative $dM/dB$ across the $\mathcal{T}_+$ transition.}
		\label{fig::derivative}
	\end{center}
\end{figure}

Based on these measurements a temperature -- magnetic field phase diagram of \CCO can be constructed, as presented in Figure~\ref{fig::phasediagram}. The zero-field region is characterized by a complicated magnetic structure, which is suggested to be a temperature- and time-dependent mixture of SDW and a collinear AFM~\cite{Agrestini2011}. When magnetic field is applied with $T < 10$ K, a given configuration of $\uparrow$ and $\downarrow$ spins on a triangular lattice is locked down and plateaus are formed. Above 10 K a single \UUD configuration exists. Further increase of field leads to a transition into a fully polarized \UUU state.

\begin{figure}[t]
	\begin{center}
		\includegraphics[width=0.9\columnwidth]{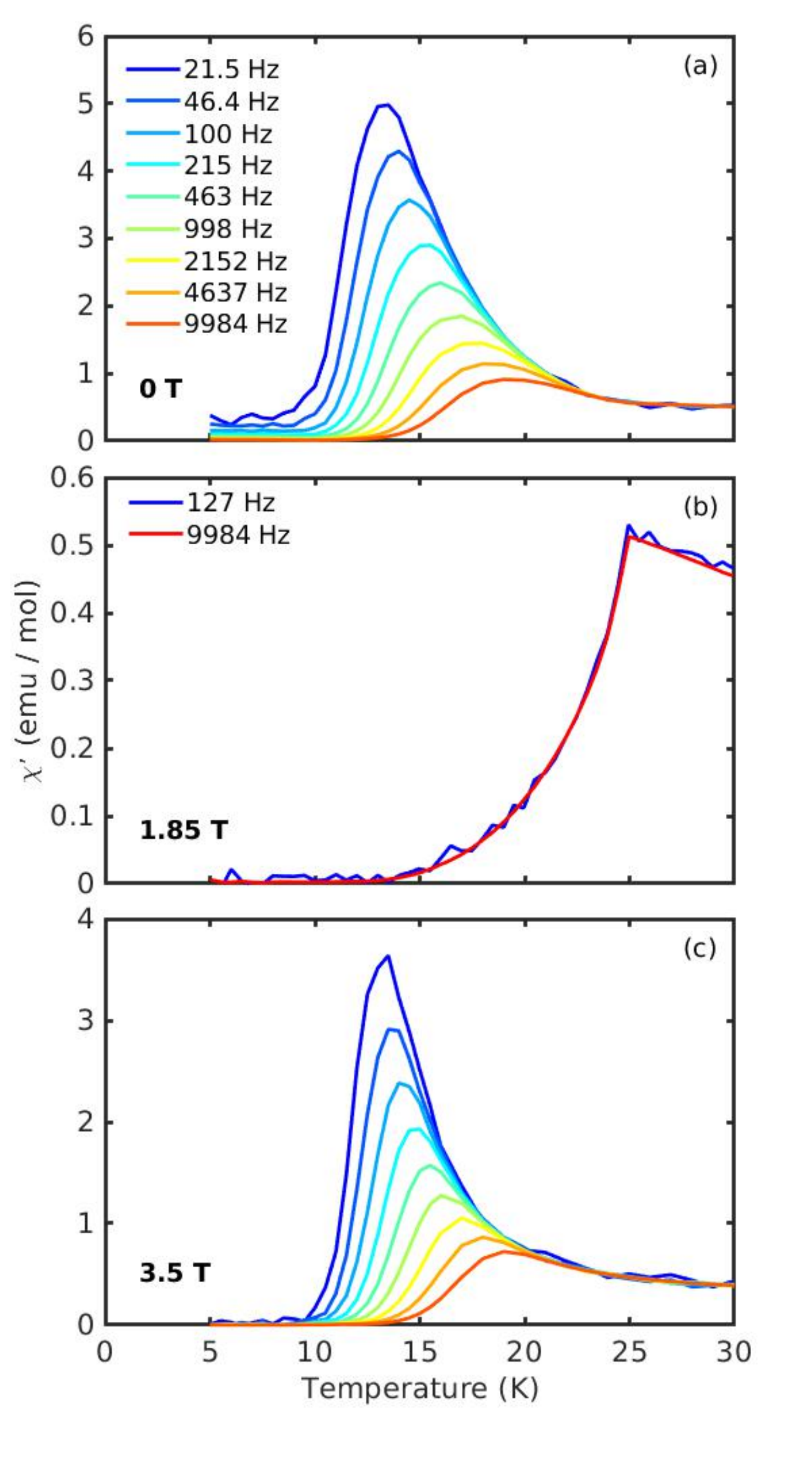}
		\caption{Temperature dependence of ac susceptibility in \CCO for (a) 0 T, (b) 1.85 T and (c) 3.5 T. The list of frequencies in (a) is the same in (c).}
		\label{fig::Tdependence}
	\end{center}
\end{figure}

The dynamic features of this system are revealed by ac magnetic susceptibility, containing two components, in-phase \real and out-of-phase \imag, often referred to as a real and imaginary component, respectively. In the low frequency limit \real is often associated with the derivative of magnetization with respect to magnetic field $dM/dB$ while \imag is related to dissipative processes during a single cycle of magnetic field. We start with the temperature dependence at three characteristic fields: (a) $B = 0$ (the middle of the $\mathcal{T}_0$ transition), (b) $B = 1.85$ T (the $1/3M_s$ plateau) and (c) $B =3.5$ T (the middle of the $\mathcal{T}_+$ transition) in Figure~\ref{fig::Tdependence}.

At zero field the transition into LRO sets in at $T_N$, indicated by neutron diffraction~\cite{Petrenko2005} and specific heat results~\cite{Hardy2003}. In ac susceptibility, however, the transition leaves practically no signature, as all the curves in Figure~\ref{fig::Tdependence}a are feature-less across $T_N$. The maximum in \real that develops below $T_N$ is strongly frequency dependent ($\Delta T_M/(T_M \Delta \omega) \sim 0.15$), mimicking a behavior observed in superparamagnetic systems~\cite{Mydosh1993}.

Around the middle of the $1/3M_s$ plateau the transition is clearly seen as a kink in susceptibility below which the signal drops to a (practically) zero value, see Figure~\ref{fig::Tdependence}b. Contrary to the highly dynamic case at $B = 0$, there is no measurable frequency dependence, indicating a robust \UUD state.

\begin{figure*}
	\begin{center}
		\includegraphics[width=1.95\columnwidth]{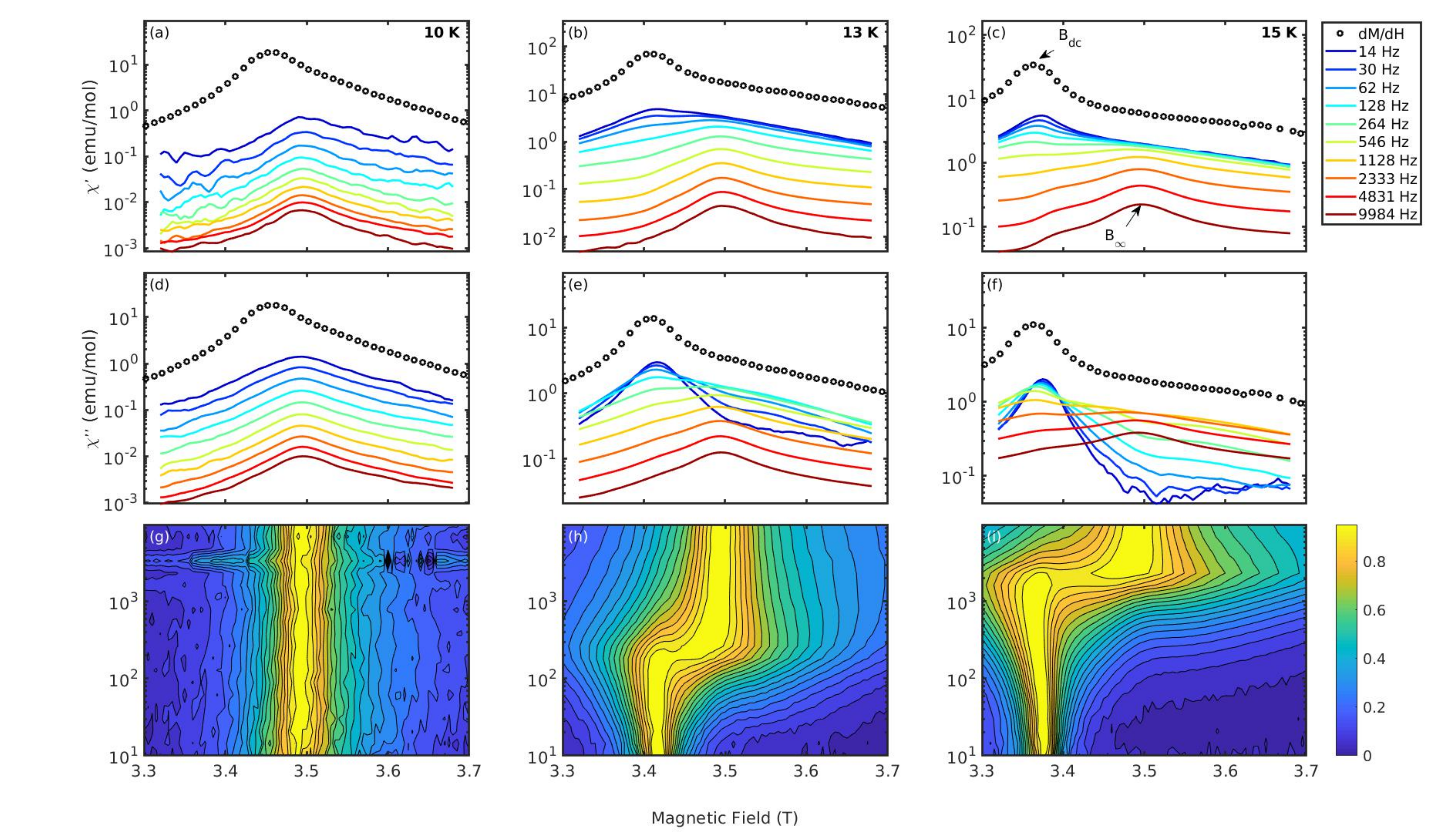}
		\caption{AC susceptibility across the transition: (a) -- (c) real component, (d) -- (f) imaginary component, (g) -- (i) colormaps of imaginary component where the maximum amplitude of each frequency profile is normalized to unity. Panels (a) -- (f) are smoothed using a moving average.}
		\label{fig::ACFieldFreq}
	\end{center}
\end{figure*}

Further increasing magnetic field to the middle of the $\mathcal{T}_+$ transition reveals a recurrence of frequency-dependent dynamics, as shown in Figure~\ref{fig::Tdependence}c. Compared to the zero field case the width of the dynamic region is somewhat narrower for each frequency and the amplitude of \real is a bit reduced. Additionally for $T < 10$ K the signal is practically zero for all frequencies, in contrast to the presence of significant dynamics at low frequencies in Figure~\ref{fig::Tdependence}a. Nevertheless, the overall appearance is very similar to the zero-field case and provides credence to a common framework of dynamics in \CCO.

We focus our attention to $T > 10$ K to avoid the issues of hysteresis and strong time- and temperature dependences. As presented in Figure~\ref{fig::derivative}, the maximum value of $dM/dB$ shows a strong decrease of its value below 10 K, indicating an emergence of a different magnetic behavior for $T < 10$ K, similar to what has been deduced for the zero-field transition.

Ac susceptibility behaviour across the $\mathcal{T}_+$ transition at several temperatures is shown in Figure~\ref{fig::ACFieldFreq}, with a corresponding $dM/dB$ included for comparison. $dM/dB$ exhibits a single maximum at $B_{dc}$, shifting to lower fields with increasing temperature. This corresponds to a shift of the steep increase of magnetization observed in Figure~\ref{fig::Hdependence}. The behavior of ac susceptibility is more complex and shows a peculiar frequency dependence, with two characteristic features emerging. The first feature is temperature dependent and it follows the same trend as $B_{dc}$, indicating that it reflects a behavior characteristic for low frequencies. The second feature is temperature independent and centered at $B_{\infty} = 3.49$ T. The observed behavior is independent of the sweep direction of magnetic field.

It is worth noting that both \real and \imag demonstrate qualitatively the same behavior but there are some qualitative differences that can be observed. At all temperatures the $B_{\infty}$ peak is very wide, spread out across the transition. On the other hand the $B_{dc}$ peak associated with the dc regime seems to exhibit smaller width and gets more pronounced in the imaginary component.

In order to better emphasize the temperature, magnetic field and frequency dependence of those two features the imaginary component in the form of color-plots is presented in Figure~\ref{fig::ACFieldFreq}g-i. Taking into account several orders of magnitude difference in the values of susceptibility measured across the frequency range used in this study, each field scan is normalized at a given frequency to its maximum value. This allows the observation of several important aspects of the $\mathcal{T}_+$ transition. With two well-defined features, there is a cross-over region in frequency where both of them are present. At high temperatures, where the separation is substantial compared to the width of peaks, there are indeed two maxima in \imag at a given frequency (for example $\nu = 2333$ Hz at $T = 15$ K). At lower temperatures the cross-over shifts to lower frequencies and the separation between the two features becomes smaller, making it harder to distinguish in field scans for a single frequency. Finally, approaching $T = 10$ K the cross-over frequency drops below 10 Hz (the lower limit of our measurements) while at the same time it becomes practically impossible to distinguish $B_{dc}$ and $B_{\infty}$.

\begin{figure}[t]
	\begin{center}
		\includegraphics[width=0.9\columnwidth]{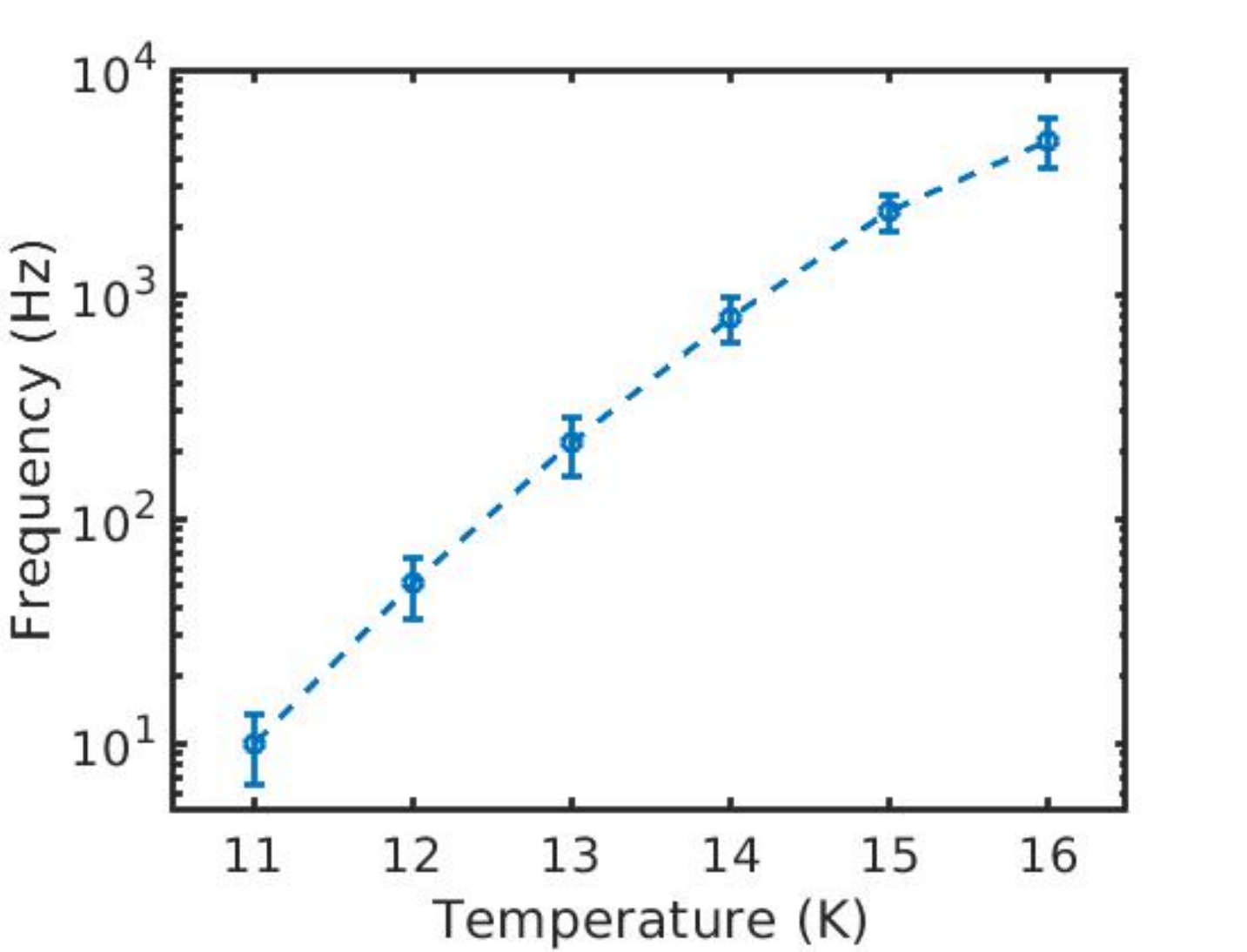}
		\caption{Temperature dependence of the cross-over frequency.}
		\label{fig::crossover}
	\end{center}
\end{figure}

The temperature dependence of the cross-over frequency is plotted in Figure~\ref{fig::crossover}. It spans three orders of magnitude in a relatively narrow temperature window. At higher temperatures a weak curvature is present, indicating a possible saturation. Experimental setups allowing higher frequency range would be needed to investigate its development towards $T_N$.~\cite{Riordan2019} On the low temperature side one would also profit from the extension towards lower frequencies with a caveat that the diminishing difference between $B_{dc}$ and $B_{\infty}$ will make the determination of the cross-over frequency ambiguous.

\begin{figure}
	\begin{center}
		\includegraphics[width=.9\columnwidth]{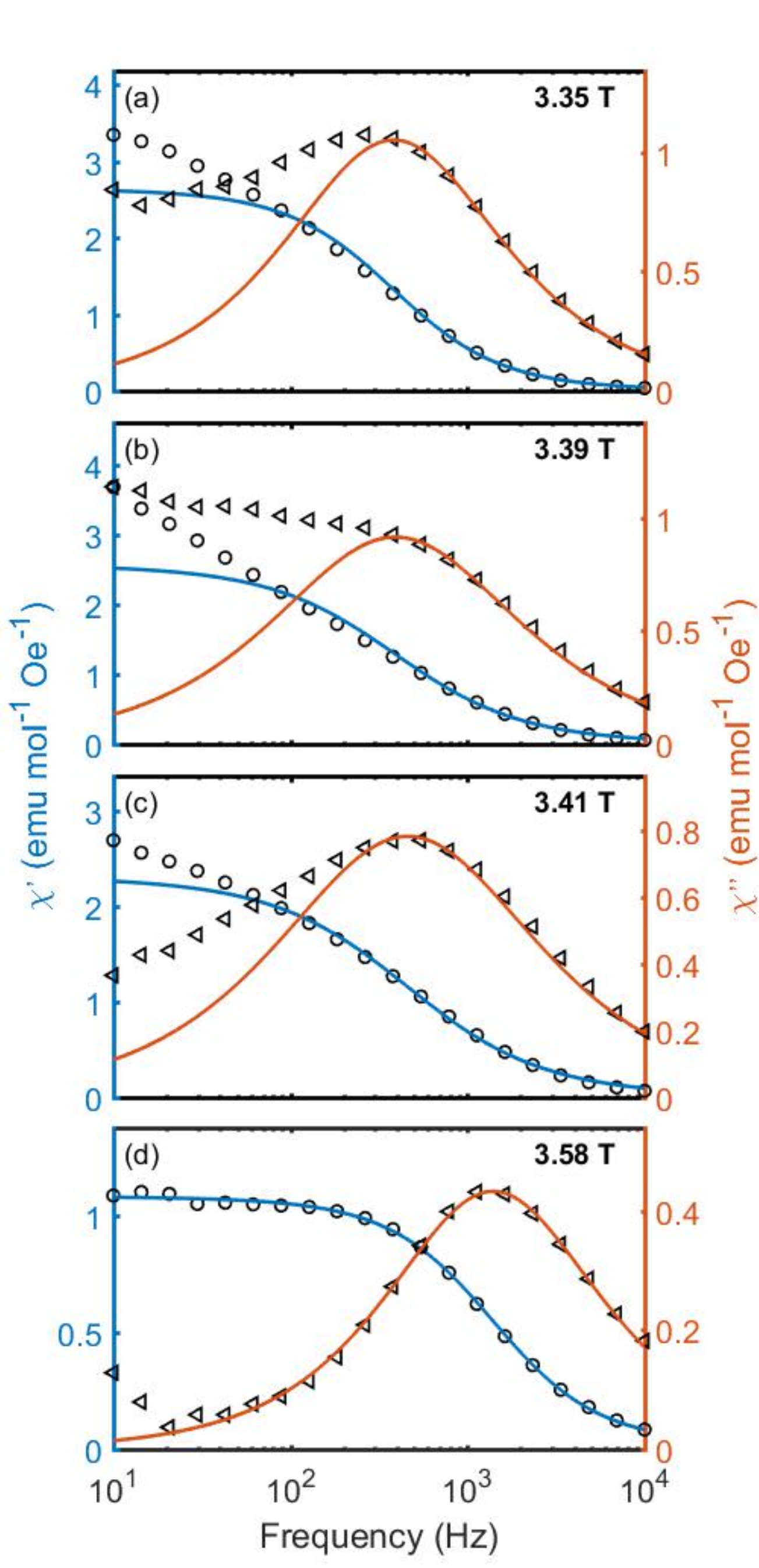}
		\caption{Frequency dependence of susceptibility at $T = 15$ K for several magnetic fields. Circles represent $\chi^{,}$, triangles $\chi^{,,}$.}
		\label{fig::Debye}
	\end{center}
\end{figure}

It is instructive to look more closely into the frequency dependence of the susceptibility. At 10 K the values of \imag monotonically increase with decreasing frequency in the whole field range investigated (Figure~\ref{fig::ACFieldFreq}b). On the other hand panels (d) and (f) show a strong non-monotonic behavior. A typical frequency dependence of both \real and \imag is plotted in Figure~\ref{fig::Debye}a, with parameters $T = 15$ K and $B = 3.58$ T. \real shows a saturation at low frequencies and diminishes at high frequencies while \imag exhibits a bell-shaped curve with a maximum occurring at the inflection point of \real. All these features are characteristic for a system whose relaxation time $\tau = 1/(2\pi f_{max})$, with $f_{max}$ being the value of frequency where the maximum in \imag occurs. Typically one finds $f_{max}$ below 10 kHz in spin-glasses and superparamagnets.~\cite{Mydosh1993} As mentioned before, the temperature dependence of the maximum in susceptibility strongly suggests superparamagnetic-like behavior in this compound, which also agrees with previous reports.~\cite{Maignan2000,Hardy2004a}

It is rarely found that only a single relaxation time is present in a system. More often there is a distribution of relaxation times. In the case of superparamagnets the distribution comes from a non-uniform size and/or shape distribution and even the environment can influence how the particles respond to magnetic field. We fit the frequency scans using the extended Debye model, where the distribution of relaxation times is phenomenologically accounted for by introducing the parameter $\alpha$ in Eq.~\ref{eq::Debye1}:

\begin{equation}
\label{eq::Debye1}
\chi (\omega) = \chi_{\infty} + \frac{\chi_{0} - \chi_{\infty}}{1 + (i \omega \tau)^{1-\alpha}}
\end{equation}

which leads to the following expressions for \real and \imag :

\begin{equation}
\label{eq::DebyeRE}
\chi^{,} (\omega) = \chi_{\infty} + \frac{1}{2}(\chi_{0} - \chi_{\infty})[1 - \frac{\rm sinh((1-\alpha)\textbf{x})}{\rm cosh((1-\alpha)\textbf{x}) + \rm sin(\alpha \pi / 2)}]
\end{equation}
\begin{equation}
\label{eq::DebyeIM}
\chi^{,,} (\omega) = \frac{1}{2}(\chi_{0} - \chi_{\infty})[\frac{\rm cos(\alpha \pi / 2)}{\rm cosh((1-\alpha)\textbf{x}) + \rm sin(\alpha \pi / 2)}]
\end{equation}

where $\textbf{x}$ = $\rm ln(\omega\tau)$, $\chi_{0}$ and $\chi_{\infty}$ are susceptibilities in the zero and infinite limits, respectively, and $\alpha = 0$ for a single characteristic time.

The applicability of this phenomenological approach to measured data is presented in Figure~\ref{fig::Debye} for several characteristic fields at $T = 15$ K. Above $B_{\infty}$ the whole of our frequency range can be rather well described with equations Eq.~\ref{eq::DebyeRE} and~\ref{eq::DebyeIM} (Figure~\ref{fig::Debye}d). Below $B_{\infty}$ a systematic increase of the low frequency region can be observed, resulting in an asymmetric frequency profile of the imaginary component (Figure~\ref{fig::Debye}c). In this case we opted to use only the high frequency side for the fit. With further decrease of magnetic field and approaching $B_{dc}$ the low frequency side becomes dominant and the former maximum becomes a pronounced shoulder (Figure~\ref{fig::Debye}b). For these values of magnetic field the determination of the high frequency region used for the fit becomes somewhat ambiguous and the extracted values should be considered only as an estimate. On the other side of $B_{dc}$ the low frequency contribution reduces again, with a maximum in imaginary component reappearing in the same frequency window around 400 Hz (Figure~\ref{fig::Debye}a).

\begin{figure}
	\begin{center}
		\includegraphics[width=.9\columnwidth]{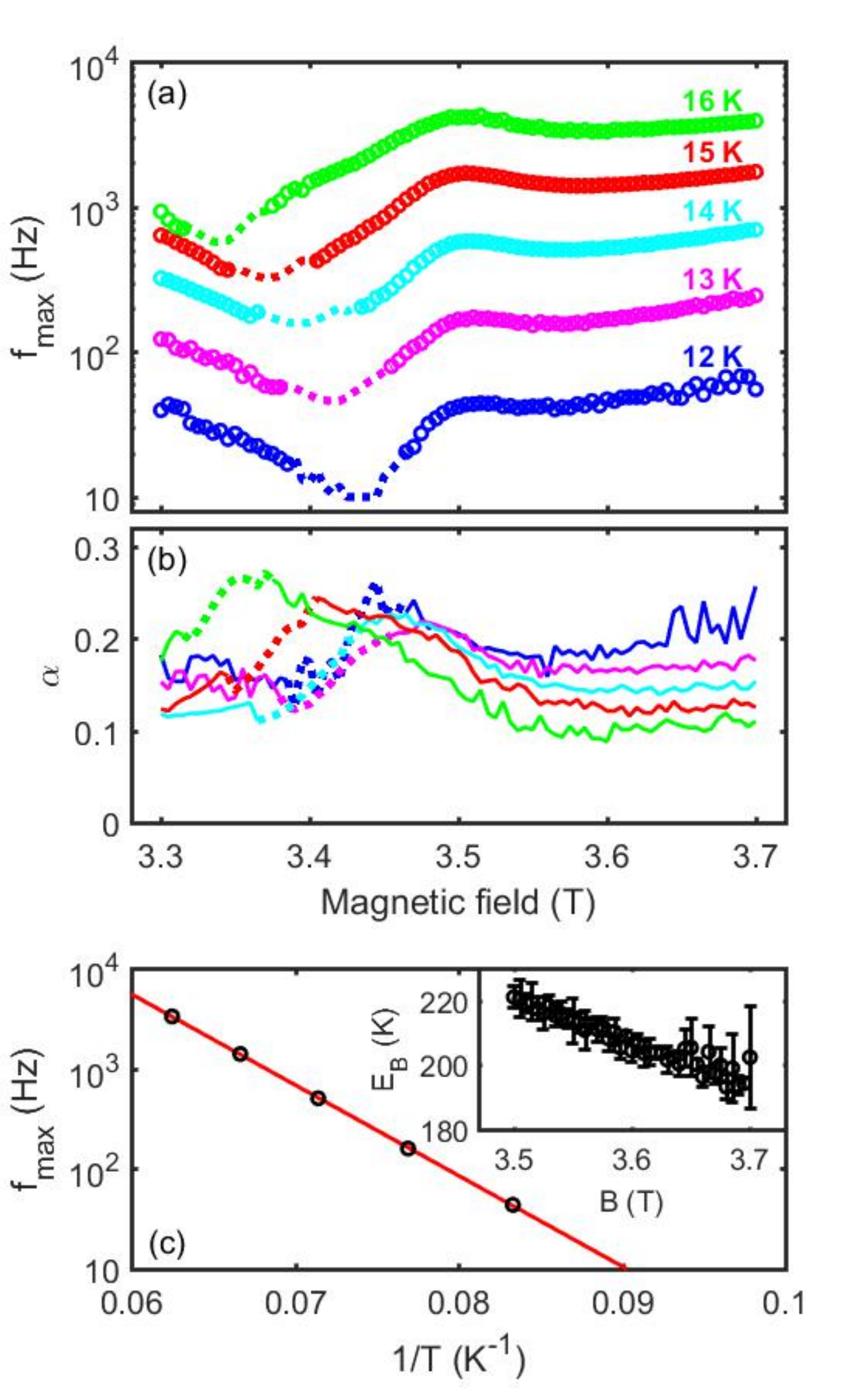}
		\caption{(a) Magnetic field and temperature dependence of the frequency $f_{max}$ where $\chi^{,,}$ exhibits a maximum. (b) The same dependence for the width of individual frequency profiles. (c) Temperature dependence of the characteristic frequency at $B = 3.58$ T. The solid line is a fit following the Arrhenius law described in Eq.~\ref{eq::Arrhenius}. The inset shows $E_B(B)$ above $B_{\infty}$.}
		\label{fig::DebyeResults}
	\end{center}
\end{figure}

The temperature and magnetic field dependence of the extracted parameters can therefore be studied. In Figures~\ref{fig::DebyeResults}a-b the magnetic field dependence of $f_{max}$ and $\alpha$ is presented, respectively, for several temperatures. Alongside the overall shift in frequency with temperature, one can observe that all the curves follow a rather uniform pattern. For $B > B_{\infty}$ there is a relatively constant $f_{max}$, with a slight curvature that at higher temperatures produces a shallow minimum around 3.6 T. At the same time $\alpha$ values are found in the range 0.1 -- 0.25, considerably lower than the value found in the similar analysis at zero field and $T = 2$ K ($\alpha \sim 0.55$ in Ref.~\onlinecite{Hardy2004a}).


For fields below $B_{\infty}$ there is an obvious trend of decreasing $f_{max}$ with a close to linear relationship between $log(f_{max})$ and $B$. Within the region around $B_{dc}$ (marked by a thick dotted line) a minimum of $f_{max}$ is found below which the trend reverses.

In the same magnetic field range the value of $\alpha$ remains around 0.25 or below. There is a peculiar inversion with respect to $B > B_{\infty}$, $T = 16$ K values have increased and $T = 13$ K have decreased. $T = 12$ K behavior stands out, which can be attributed to the fact that the values of $f_{max}$ are rather low and not easily separable from the observed low frequency upturn. One should anyway refrain from putting too much emphasis on the values of $\alpha$ when they are extracted using only half of the frequency profile. To get a reliable set one would need to include the low frequency upturn into the analysis for which a significantly lower frequency range is required.

Given the uncertainty in determination of the exact set of parameters in the region around the minimum, we plot them as thick dotted lines, compared to the rest of the data set where symbols (for $f_{max}$) and full lines (for $\alpha$) are used. Nevertheless, it is obvious that there must be a minimum in $f_{max}$ vs $B$ since on both sides there is a reliable set of data which shows a (quasi-)linear behaviour with opposite slopes. Additionally, the region with a dominant low frequency contribution (Figure~\ref{fig::Debye}b) is rather narrow (except at $T = 12$ K) so we do not expect that different approaches would result in vastly different values of minima. The approach that we used results in a minimum of $f_{max}$ at $B = 3.37$ T while $B_{dc} = 3.36$ T for $T = 15$ K, strongly suggesting that the two features have common underlying processes.

A relatively uniform behavior of $f_{max}$ above $B_{dc}$ allows the quantification of its temperature dependence. Figure~\ref{fig::DebyeResults}c shows the extracted data at $B = 3.58$ T, together with a fit to a simple activation energy dependence:

\begin{equation}
\label{eq::Arrhenius}
f_{max} = f_0 e^{-\frac{E_B}{k_B T}}
\end{equation}

where $f_0$ is an attempt frequency, $E_B$ is an energy barrier that a correlated cluster of moments needs to overcome to flip its direction, and $k_B$ is the Boltzmann constant. In the inset of the same panel we plot $E_B(B)$ where it is shown that there is a very weak magnetic field dependence, with values remaining around 190 -- 220 K. For comparison, previously extracted value at $T = 2$ K, $B = 0$ is 135 K.~\cite{Hardy2004a} The values of $f_0$ are found to be at the order of $10^9$ Hz.

\section{Discussion}

The intricacies of the ground state~\cite{Maignan2000}, extremely long time scales associated with various magnetic configurations~\cite{Agrestini2011} and puzzling dynamics~\cite{Hardy2004a} have kept the focus of the research on \CCO for long time onto the $B = 0$ case. The quest was to find the equilibrium state, as is a common approach for majority of magnetic materials. On the other hand, in recent years a growing interest of the research community is directed towards understanding the non-equilibrium properties, their governing principles and applicability to individual systems.~\cite{Jarzynski2015} As emphasized in the introduction, the $B = 0$ state of \CCO is a peculiar point in the middle of the $\mathcal{T}_0$ transition, so to truly understand it, the processes that control its field-induced non-equilibrium evolution need to be understood.

By focusing on the $\mathcal{T}_+$ transition above 10 K we avoid the complications that arise with exceptionally long time scales and hysteretic behavior. Within the $\mathcal{T}_+$ transition a (small) change of magnetic field $\Delta B$ causes a certain number of spins to change the orientation, inducing $\Delta M$. There is an energy barrier associated with this spin reorientation, $E_B$, allowing us to investigate the associated time scales through the Arrhenius law. Indeed, above 3.5 T, where the majority of spins are pointing up, the flipping cluster of down spins seems to be well defined, with the energy barrier practically independent of magnetic field. The values of $\alpha$, extracted from individual fits of frequency dependence (Figure~\ref{fig::DebyeResults}c), are found between 0.1 and 0.2 for magnetic fields above 3.5 T. These values can be compared to the value extracted using the same approach at $T = 2$ K and $B = 0$ in Ref.~\onlinecite{Hardy2004a} ($\alpha \sim 0.55$). More generally, in systems where a distribution of time scales dominates the dynamics, like spin-glasses and superparamagnets, the values of $\alpha$ are significantly larger. It has been found~\cite{Huser1986} that in $Cu$Mn (5 at. \%) and (Eu$_{0.4}$Sr$_{0.6}$)S, examples of archetypal spin-glasses, $\alpha$ linearly drops from 1 (an infinite width) at the lowest temperature to 0.2 around $2T_f$ ($T_f$ being the freezing temperature of a spin-glass system). In the same report they compare this to two superparamagnets where the value of $\alpha$ accumulates between 0.5 and 0.7, independent of temperature. Looking at Figure~\ref{fig::DebyeResults}c and observing the temperature dependence of $\alpha$ in \CCO above 3.5 T it appears that at temperatures around 20 K one would enter a regime where only a single relaxation time is found ($\alpha = 0$). If such a prediction appears to be correct, this would open up a possibility to utilize microscopic methods to characterize and eventually understand the smallest dynamic unit in \CCO.

The value of the energy barrier has been determined to be somewhat larger than previously reported~\cite{Hardy2004a}, albeit at the different temperature and magnetic field. On the other hand it has been determined from inelastic neutron scattering that a large spin gap of $\sim 27$ meV ($= 313$ K) characterizes the magnon dispersion in \CCO, which has been attributed to the large single-ion anisotropy.~\cite{Jain2013} If one follows the idea put forward through Monte Carlo simulations~\cite{Kamiya2012}, a simple flipping process would be ascribed to a soliton being moved along the chain direction. A soliton is a domain wall boundary between uniformly magnetized chain segments and its lattice has been determined to be entropy driven, giving rise to incommensurate diffraction peaks in the SDW phase.~\cite{Agrestini2008}

Below 3.5 T a qualitatively different behavior is seen, with a second energy scale appearing at low frequencies. Around $B_{dc}$ this contribution becomes dominant and for a proper qualitative analysis much lower frequencies are needed. On the other hand the field dependence of $f_{max}$ can still be traced, remaining visible as a shoulder in individual frequency scans and developing a minimum at $B_{dc}$. The existence of two energy scales could be associated with intra-chain and inter-chain correlations, the later forming large, percolating clusters that have their characteristic frequencies much lower than the intra-chain ones. Due to the weak coupling between the chains, an individual fingerprint of intra-chain correlated cluster at higher frequencies can still be seen.

At the moment it is not clear what kind of processes govern the appearance of the temperature-independent dissipation maximum at $B_{\infty}$. Given that $\chi_{0}$ from Eq.~\ref{eq::Debye1} also exhibits a maximum at this field (not shown) we can speculate that $B_{\infty}$ simply reflects the maximum in the number of driven clusters. However, the peculiar frequency dependence, with a cross-over to a 'static regime' marked by $B_{dc}$ remains puzzling. In that context, it would be interesting to extend the observation of the cross-over frequency above our 10 kHz limit, where the collected data (Figure~\ref{fig::crossover}) do indicate a possible saturation at higher temperatures. Additionally, both $B = 0$ and $B = 3.5$ T susceptibility is rather impervious to the appearance of LRO at $T_N$ (Figure~\ref{fig::Tdependence}), indicating that the dynamic part extends into the paramagnetic regime. This is in principle not surprising if the core of magnetic clusters is made out of single-chain segments which show intra-chain correlations well above $T_N$.~\cite{Takeshita2006,Jain2013,Mansson2018}

At the end we should mention that the study of the $\mathcal{T}_+$ transition in \CCO extends beyond the immediate relevance for this compound. Recently several new materials emerged that show similar plateaus and jumps in magnetization as \CCO. First there are two polymorphs of CoV$_2$O$_6$, possessing a similar albeit distorted triangular lattice in the $ab$ plane, thus lifting the frustration and exhibiting a typical AFM spin arrangement.~\cite{Lenertz2011} Another interesting compound is CoCl$_2 \cdot$2H$_2$O, which orders as a two sublattice antiferromagnet~\cite{Mollymoto1980} and has been studied as function of transverse fields.~\cite{Larsen2017} Although their ground states differ substantially from the one found in \CCO (and between each other), they do share the existence of the $1/3M_s$ and $M_s$ plateaus, with a very similar, temperature-induced asymmetric magnetization profile as found for $\mathcal{T}_+$ in \CCO. Further insights into their dynamics across this transition would allow to separate the material-specific from the more general features ascribed to non-equilibrium processes.

\section{Summary}

We have presented a detailed study of the dynamic aspects of the transition between two plateaus in magnetization of \CCO around 3.5 T. It has been revealed that the region between the saturation and the middle of the transition can be described within a framework of relatively narrow distribution of cluster sizes. The middle of the transition, characterized by the fastest change in magnetization, exhibits a complicated frequency response, with two characteristic magnetic fields that correspond to low and high frequency regimes, respectively. Individual frequency scans also demonstrate the presence of two characteristic time scales, which we have tentatively ascribed to intra-chain and inter-chain correlations.

\section{Acknowledgements}
The work was supported in part by the Swiss National Science Foundation through grants No.200021-169699, No.200020-188648 and No.206021-189644.


%

\end{document}